\documentclass[showpacs,twocolumn,showkeys,amsmath,amssymb,pra]{revtex4-1}

\usepackage{bm}
\usepackage{bbm}
\usepackage{dsfont}
\usepackage{color}
\usepackage{graphicx}   
\usepackage[colorlinks=true,urlcolor=blue,linkcolor=green]{hyperref}

\newcommand{\be}{\begin{equation}}
\newcommand{\ee}{\end{equation}}
\newcommand{\rd}{{\mathrm d}}
\newcommand{\re}{{\mathrm e}}
\newcommand{\ri}{{\mathrm i}}

\newcommand{\ab}{a^{\phantom{\dagger}}}
\newcommand{\ad}{a^{\dagger}}
\begin{document}

\title[Degree of coherence]
      {Degree of simplicity of Floquet states of a periodically driven
	Bose-Hubbard dimer}  

\author{Steffen Seligmann and Martin Holthaus}
%{Contact author: martin.holthaus@uol.de} 
		
\affiliation{Institut f\"ur Physik, Carl von Ossietzky Universit\"at,
	D-26111 Oldenburg, Germany}	
                  
\date{July 14, 2025}

\begin{abstract}
	We investigate numerically computed Floquet states of a Bose-Hubbard 
	dimer that is subjected to strong, time-periodic forcing with respect 
	to their coherence, invoking a measure for their degree of simplicity 
	previously suggested by Leggett. This serves to ascertain the validity 
	of the mean-field approximation under conditions such that the 
	time-dependent nonlinear Gross-Pitaevskii equation has chaotic 
	solutions. It is shown that for sufficiently large particle numbers 
	the exact $N$-particle Floquet state semiclassically associated with 
	the innermost quantized invariant tube surrounding a stable periodic 
	mean-field orbit represents a macroscopically occupied single-particle 
	state, {\em i.e.\/}, a Floquet condensate. 
\end{abstract} 


\maketitle 

%%%%%%%%%%%%%%%%%%%%%%%%%%%%%%%%%%%%%%%%%%%%%%%%%%%%%%%%%%%%%%%%%%%%%%%%%%%%%%%%

\section{Introduction}
\label{S_1}

The physics of quantum systems that are driven peri\-odically in time has 
spurred a host of activities recently~\cite{Eckardt17,BukovEtAl15,
GoldmanDalibard14,Holthaus16}, additionally fueled by the celebrated 
Floquet time crystals~\cite{KhemaniEtAl16,ElseEtAl16,ZhangEtAl17,ChoiEtAl17,
PizziEtAl19,KhemaniEtAl19,ElseEtAl20,ZalatelEtAl23}. In the present study we 
utilize the Floquet picture in order to investigate the question under which 
circumstances there might be Floquet condensates, representing macroscopically 
occupied single-particle Floquet states of a periodically driven 
non-equilibrium system comprising $N$ interacting Bose particles. 
 Indeed, the very existence of Bose-Einstein condensates does not
 necessarily require that a many-body Bose system be in thermal equilibrium,
 nor even in a steady state, as has previously been emphasized by 
 Leggett~\cite{Leggett01}. Along these lines, condensation in 
 driven-dissipative ideal Bose gases has been investigated on the basis 
 of rate equations~\cite{SchnellEtAl18}, with application to open driven 
 optical-lattice systems~\cite{SchnellEtAl23}, while the preparation of  
 Floquet condensates in interacting, periodically driven, but otherwise 
 isolated systems by means of an adiabatic turn-on of the drive had been 
 envisioned in Ref.~\cite{HeinischHolthaus16}. While we do not tackle the 
 pivotal question concerning the possible emergence of Floquet condensates 
 in its full generality here, we resort to an idealized model that allows one 
 to check each supposition by exact numerical computations, namely, a 
 periodically driven Bose-Hubbard dimer, and thereby to provide an affirmative 
 answer. We will introduce the model in Sec.~\ref{S_2} and 
 report the results of  our analysis in Sec.~{\ref{S_3}}, addressing, among 
 others, the validity of the time-dependent Gross-Pitaevskii equation under 
 conditions such that the mean-field dynamics are partly chaotic. 
 This enables one, in particular, to transfer insight previously 
 gained in the study of the correspondence between chaotic classical 
 single-particle systems and their quantum mechanical 
 counterparts~\cite{Gutzwiller90,Haake10}  to the correspondence between 
 the approximate mean-field description and the full quantum dynamics of  
 $N$-particle Bose systems, with the semiclassical limit $\hbar \to 0$ being 
 replaced by the hypothetical limit $N \to \infty$.  In this way, one is led to 
 a pertinent deduction: While the anti\-cipated Floquet condensates would be 
 permanently stable within a mean-field approximation, they could eventually 
 be rendered metastable due to a subtle beyond-mean-field quantum effect, 
 although such metastability might not manifest itself on experimentally 
 relevant time scales.     

In the course of our discussion, we also provide some background 
information on basic elements of the Floquet approach and the semiclasscial 
quantization of perio\-dically driven systems. This should not only serve to 
establish our notation, but also help to make the material accessible to 
nonspecialists without undue hardship. Section~\ref{S_4} then 
briefly sums up our main assertions, including a suggestion for 
actual  laboratory verifications of our model-based tentative predictions.

\section{The periodically driven Bose-Hubbard dimer}
\label{S_2}

The Bose-Hubbard dimer constitutes a minimalistic model of quantum many-body
physics. It describes~$N$ Bose particles that occupy two sites, are allowed to
tunnel from one site to the other, and are endowed with an on-site interaction 
that we take to be repulsive here. Denoting the strength of the tunneling 
contact by $\hbar\Omega$ and the interaction strength by $\hbar\kappa$, we 
write its Hamiltonian in the form 
\begin{eqnarray}
	H_0 & = & -\frac{\hbar\Omega}{2} 
	\left( \ad_2\ab_1 + \ad_1\ab_2 \right) 
\nonumber \\ & & 
	+ \hbar\kappa \left( \ad_1\ad_1\ab_1\ab_1 
	+ \ad_2\ad_2\ab_2\ab_2 \right) \; , 	
\label{eq:BHD}
\end{eqnarray}
where $\ab_j$ and $\ad_j$ are, respectively, the annihilation and creation
operators referring to the site labeled~$j$, obeying the usual bosonic 
commutation relations $(j,k = 1,2)$:
\be
	\left[ \ab_j, \ab_k \right] = 0 	\; , \quad 
	\left[ \ad_j, \ad_k \right] = 0 	\; , \quad
	\left[ \ab_j, \ad_k \right] = \delta_{jk} \; .  
\ee	
This system~(\ref{eq:BHD}) serves as a simplistic model for a Bose-Einstein
condensate in a double-well potential~\cite{MilburnEtAl97,ParkinsWalls98,
Leggett01} and has been applied for a long time and by many authors to an
abundant variety of subjects such as decoherence~\cite{VardiAnglin01}, 
multiple-timescale  dynamics~\cite{KalosakasEtAl03}, 
quantum phase-space analysis~\cite{MahmudEtAl05}, 
phase-diffusion dynamics~\cite{BoukobzaEtAl09},
finite-rate quenches~\cite{VenumadhavEtAl10}, 
WKB quantization~\cite{SimonStrunz12}, 
density functional theory~\cite{CarrascalEtAl15},
thermalization~\cite{KiddEtAl20}, 
and entanglement generation~\cite{SolankiEtAl25},
to name but a few. Most notably, there exists a faithful experimental 
realization by means of ultracold atoms in an optically generated double well, 
referred to as a bosonic Josephson junction~\cite{GatiOberthaler07}.
For computational purposes, one of its benefits lies in the fact that the 
$N$-particle  sector ${\cal H}_N$ of its full Fock space has the dimension
${\rm dim} \, {\cal H}_N = N+1$ that grows merely linearly with~$N$, thus
allowing one to reach large~$N$ numerically without approximations.    

 Now let us add a time-periodic driving force to this model, such that the
 on-site energies are sinusoidally modulated in phase opposition to each 
 other with angular frequency~$\omega$ and amplitude~$\hbar\mu$. The total
 Hamiltonian then reads~\cite{HolthausStenholm01,WeissTeichmann08,
 GertjerenkenHolthaus14}      
\be
	H(t) = H_0 +  H_1(t) \; , 
\label{eq:PDD}
\ee	
where
\be
	H_1(t) = \hbar\mu \sin(\omega t)
	\left( \ad_1\ab_1 - \ad_2\ab_2 \right) \; .  	
\label{eq:HDR}
\ee
Being periodically time-dependent, $H(t) = H(t+T)$, with $T = 2\pi/\omega$, the
method of choice for the analysis of the extended model~(\ref{eq:PDD}) rests on
the quantum mechanical Floquet picture~\cite{AutlerTownes55,Shirley65,Sambe73,
Salzman74,BaroneEtAl77,FainshteinEtAl78,MilfeldWyatt83}. Briefly, when 
expressing the eigenvalues of the unitary one-cycle evolution operator~$U(T,0)$ 
within ${\cal H}_N$ as $\exp(-\ri\gamma_n)$ with real phases~$\gamma_n$, and 
de\-noting the associated eigenvectors by $|n\rangle$, one has the spectral 
decomposition
\be
	U(T,0) = \sum_{n=1}^{N+1} 
	| n \rangle \, \re^{-\ri\gamma_n} \langle n |  \; .
\label{eq:DIA}	
\ee
Note that the eigenvalues tend to fill the unit circle densely when the particle 
number~$N$ becomes very large.  This may necessitate some caution in numerical
calculations, since closely spaced eigenvalues cannot always be distinguished 
by  finite-precision arithmetics. Writing the eigenphases in the suggestive
form $\gamma_n = \varepsilon_n T/\hbar$, this expansion~(\ref{eq:DIA}) is
strongly reminiscent of  the evolution of energy eigenstates of a 
time-independent Hamiltonian. Hence, the quantities $\varepsilon_n$ extracted
from the eigenphases have aptly been termed quasienergies~\cite{Zeldovich67,
Ritus67}. Note, however, that each quasienergy is thus defined only up to a 
positive or negative integer multiple of $\hbar\omega$, since the 
phases~$\gamma_n$ are defined only up  to an integer multiple of $2\pi$. 
Hence, a quasienergy should not be regarded as a single number, but rather as 
a set of equivalent representatives spaced by $\hbar\omega$. This implies that 
the quasienergy spectrum is unbounded from both above and below even for the 
model~(\ref{eq:PDD}), although ${\cal H}_N$ is of finite dimension, so that 
the quasienergies cannot be  ordered with respect to their apparent magnitude. 
This observation will be taken up again in the following section.

Extending the above stroboscopic viewpoint, the full Floquet states
$|\psi_n(t)\rangle$ that actually solve the time-dependent Schr\"odinger
equation are obtained by following the eigenvectors $| n \rangle$ 
continuously in time: 
\begin{eqnarray}
	| \psi_n(t) \rangle & = & U(t,0) | n \rangle
\nonumber \\	& \equiv &
	| u_n (t) \rangle \exp(-\ri\varepsilon_n t/\hbar) \; , 	
\label{eq:FFS}	
\end{eqnarray}
where the Floquet functions $|u_n(t)\rangle =  |u_n(t+T)\rangle$ inherit the 
periodic time-dependence of the Hamiltonian. While a stroboscopic treatment 
based on the diagonalization~(\ref{eq:DIA}) often suffices for practical 
numerical calculations, this extended approach commonly is adopted for 
mathematical and conceptual considerations~\cite{Howland89,Joye94}.

Although the numerical solution  of  the time-dependent Schr\"odinger equation 
governed by the Hamiltonian~(\ref{eq:PDD}) poses no problem even for fairly 
large~$N$, valuable insight is gained by juxtaposing the exact many-body 
dynamics to their mean-field approximation. Leaving aside formal complications
arising  from the fact that the individual annihilation and creation operators 
$\ab_j$ and $\ad_j$ remain mathematically undefined when restricted to the 
$N$-particle sector ${\cal H}_N$, so that the mean-field factorization of 
expectation values of products of such operators into products of expectation 
values requires some careful analysis when the number~$N$ of particles  is 
conserved~\cite{GertjerenkenHolthaus15}, this is effectively achieved by 
transforming  $\ab_j$ and $\ad_j$ to the Heisenberg picture, obtaining their 
time-dependent equivalents $\widetilde{a}^{\phantom\dagger}_j$ and 
$\widetilde{a}^\dagger_j$, and then replacing the latter by $c$-number
amplitudes $c_j$ and $c_j^\ast$ according to the scheme
\be 
	\widetilde{a}^{\phantom\dagger}_j(\tau)  
	\longrightarrow \sqrt{N} c_j(\tau)\; ,
	\quad
        \widetilde{a}^\dagger_j(\tau) 
        \longrightarrow \sqrt{N} c_j^\ast(\tau) \; . 	  
\label{eq:SCA}
\ee	
From here on we employ the dimensionless time variable $\tau = \Omega t$. 
These mean-field amplitudes obey a nonlinear system of equations of the 
Gross-Pitaevskii type, namely~\cite{SmerziEtAl97,RaghavanEtAl99}:
\begin{eqnarray}
	\ri\dot{c}_1(\tau) & = & -\frac{1}{2} c_2(\tau) 
	+ 2\alpha | c_1(\tau) |^2 c_1(\tau) 
\nonumber \\ & & 	
	+ \frac{\mu}{\Omega}
	\sin\!\left( \frac{\omega}{\Omega}\tau\!\right)  c_1(\tau) \; ,
\nonumber \\	
	\ri\dot{c}_2(\tau)  & = & -\frac{1}{2} c_1(\tau) 
	+ 2\alpha | c_2(\tau) |^2 c_2(\tau) 
\nonumber \\ & & 	
	- \frac{\mu}{\Omega}
	\sin\!\left( \frac{\omega}{\Omega}\tau\!\right) c_2(\tau) \; ,
\label{eq:GPE}		
\end{eqnarray} 
where we have introduced the dimensionless mean-field parameter
\be
	\alpha  = \frac{N\kappa}{\Omega} \; .	
\label{eq:ALP}	
\ee
Hence, while the $N$-particle dynamics generated by the 
Hamiltonian~(\ref{eq:PDD}) depend on $N$ and $\kappa/\omega$ separately, 
the corresponding mean-field dynamics depend only on their 
combination~(\ref{eq:ALP}), implying that the mean-field limit is approached 
when $N$ becomes large while $\kappa/\Omega$ tends to zero, such that their 
product remains constant. 

Following common practice~\cite{SmerziEtAl97,RaghavanEtAl99}, we now 
factorize the mean-field amplitudes into absolute values and phase factors,
\be
	c_j(\tau) = | c_j (\tau) | \exp\!\left(\ri\theta_j(\tau)\right) \; ,
\ee	
and introduce the population imbalance
\be
	p = | c_1 |^2 - | c_2 |^2
\label{eq:PIM}
\ee
together with the relative phase	
\be
	\varphi = \theta_2 - \theta_1 \; .
\label{eq:REP}	
\ee	
Expressed in terms of these variables, the Gross-Pitaevskii 
system~(\ref{eq:GPE}) acquires the equivalent form 
\begin{eqnarray}
	\dot{p} & = & -\sqrt{1 - p^2} \, \sin(\varphi) \; ,
\nonumber \\
	\dot{\varphi}& = & 
	2\alpha p + \frac{p}{\sqrt{1 - p^2}} \cos(\varphi) 
	+ 2 \frac{\mu}{\Omega} 
	\sin\!\left(\frac{\omega}{\Omega} \tau\!\right) \; ,
\label{eq:EQF}	
\end{eqnarray}
which amounts to the Hamiltonian equations of motion derived from the 
classical Hamiltonian function 
\be
	H_{\rm mf}(\tau) = \alpha p^2 - \sqrt{1 - p^2}  \, \cos(\varphi)	 	
	+ 2 \frac{\mu}{\Omega}  \, p \, 
	\sin\!\left(\frac{\omega}{\Omega} \tau\!\right) \; . 	
\label{eq:HMF}
\ee
This Hamiltonian evidently describes a pendulum with angular momentum~$p$
and conjugate angle~$\varphi$, which is driven periodically in time, possesses 
a mass inversely proportional to the mean-field parameter~(\ref{eq:ALP}), and 
features a length that shortens with increasing momentum~\cite{SmerziEtAl97,
RaghavanEtAl99}.

As is well known, a periodically driven nonlinear pendulum exhibits chaotic 
dynamics, in contrast to its undriven antecessor. Seen from this angle, 
the above reformulation of the Gross-Pitaevskii system~(\ref{eq:GPE}) offers 
the advantage that an eminent body of know\-ledge gathered in the study 
of  regular and chaotic motion in classical Hamiltonian
systems~\cite{AbrahamMarsden08,LiLi92,GuckenheimerHolmes02} becomes 
immediately available for the investigation of the mean-field dynamics of the 
periodically driven Bose-Hubbard dimer, while the correspondence of this 
mean-field dynamics to the full $N$-particle quantum time evolution is covered 
by the correspondence of classical periodically driven single-particle systems 
to their quantum mechanical counterparts~\cite{Gutzwiller90,Haake10}. 

\begin{figure}[t]
\centering
\includegraphics[width=1.0\linewidth]{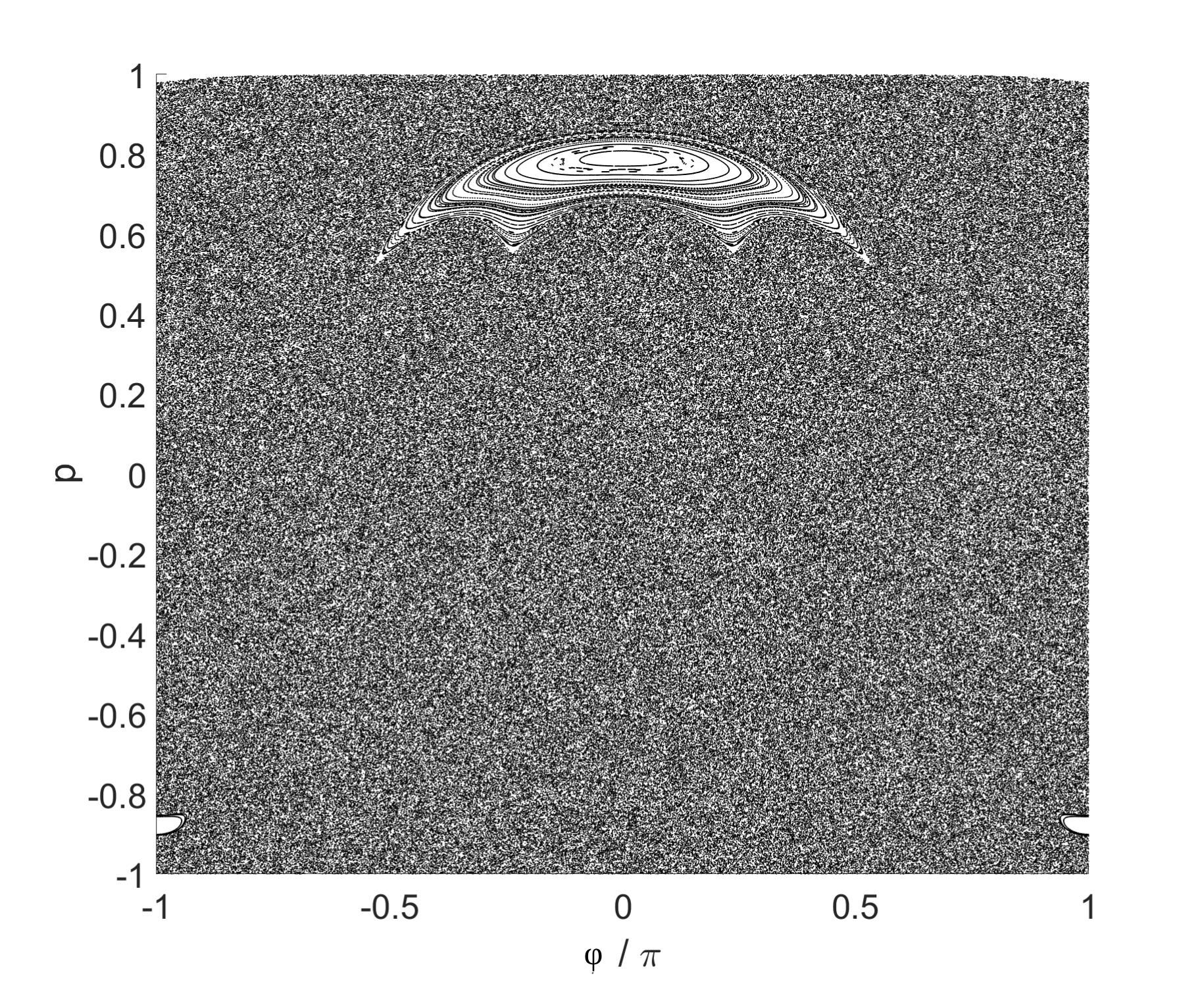}
\caption{Poincar\'e surface of section that pertains to the periodically 
driven pendulum with momentum-dependent length~(\ref{eq:HMF}), visualizing 
the mean-field dynamics of the periodically driven Bose-Hubbard dimer
$H_0 + H_1(t)$ defined by Eqs.~(\ref{eq:BHD}) and (\ref{eq:HDR}). 
Parameters here and for all following figures are $\alpha = 1.30$, 
$\mu/\Omega = 0.41$, and $\omega/\Omega = 1.40$.}  
\label{F_1}
\end{figure}

As a guiding step in this direction, we display in Fig.~\ref{F_1} a Poincar\'e 
surface of section for the periodically driven pendulum~(\ref{eq:HMF}) 
with mean-field parameter~$\alpha = 1.30$, scaled driving amplitude 
$\mu/\Omega = 0.41$, and scaled driving frequency $\omega/\Omega = 1.40$;
these parameters will be employed throughout this work~\cite{Data}. 
Such a section is produced by selecting a suitable set of initial 
values $(p_0,\varphi_0)$ in the phase-space plane, solving Hamilton's 
equations for one driving period and recording the resulting image point 
$(p_1,\varphi_1)$, and iterating, obtaining the successors $(p_k,\varphi_k)$  
after $k$~periods. The underlying Poincar\'e map from the phase-space plane 
to its image under the Hamiltonian flow after one period thus constitutes the 
classical cor\-respondent to the quantum period mapping mediated by the 
one-cycle evolution operator~(\ref{eq:DIA}). One observes in Fig.~\ref{F_1} 
the coexistence of regular and stochastic motion which is typical for 
Hamiltonian systems, with a large resonant island of mainly regular motion 
at $\varphi = 0$ embedded in an apparently chaotic sea; this island hosts a 
stable periodic orbit in its center. It should be noted, however, 
that even within the seemingly regular island there exists numerically 
unresolvable fine-scale chaotic motion, due to the dissolution of all closed 
contours for which the ratio between the pendulum's unperturbed oscillation 
frequency and the driving frequency is not sufficiently irrational, in the 
sense of a continued-fraction expansion~\cite{Gutzwiller90,LiLi92}.
There also is a further tiny island of 
stable, almost regular motion at $\varphi = \pm\pi$, akin to the familiar 
$\pi$-oscillations~\cite{RaghavanEtAl99} exhibited by the undriven 
Bose-Hubbard dimer~(\ref{eq:BHD}).  The parameters employed here
have been specifically chosen such that the boundary between mainly regular
and chaotic motion is particularly sharp. In the following section, we will 
explore the ramifications of this mixed regular and stochastic mean-field 
dynamics for the actual $N$-particle dynamics of the periodically driven 
Bose-Hubbard dimer~(\ref{eq:PDD}).

\section{Floquet states and their degree of simplicity}
\label{S_3}

The bridge between the classical Poincar\'e map and the quantum 
$N$-particle Floquet states is created by the coherent spin states 
$|\vartheta,\varphi\rangle_{\!N}$ introduced and discussed  further in 
Refs.~\cite{Radcliffe71, ArecchiEtAl72}. These  are $N$-fold occupied 
single-particle states of the two-level system furnished by the two dimer 
sites, here written as   
\be
	|\vartheta,\varphi\rangle_{\!N} = 
	\frac{1}{\sqrt{N!}} \left(A^\dagger\right)^{\!N} | vac \rangle \; ,
\ee
where $|vac\rangle$ denotes the empty-dimer vacuum state, and
\be 
	A^\dagger = \cos\!\frac{\vartheta}{2} \,\ad_1 
	+ \sin\!\frac{\vartheta}{2} \, \re^{\ri\varphi} \, \ad_2
\ee
acts as a bosonic creation operator, obeying $[ A,A^\dagger] = 1$. Hence, 
the variable~$\varphi$ appearing here is identified with the relative  phase 
defined by Eq.~(\ref{eq:REP}), while the population imbalance~(\ref{eq:PIM}) 
is given by $p = \cos^2(\vartheta/2) - \sin^2(\vartheta/2) = \cos\vartheta$. 
Therefore, an $N$-particle coherent state $ |\vartheta,\varphi\rangle_{\!N}$ 
corresponds to the point $(p\!\!=\!\!\cos\vartheta,\varphi)$ in the phase-space 
plane of the periodically driven mean-field pendulum~(\ref{eq:HMF}).    
The squared overlap
\be
	Q^{(N)}_{|\psi\rangle}(\cos\vartheta,\varphi) = 
	\big| \langle \psi | \vartheta,\varphi \rangle_{\!N} \big|^2  
\ee
thus quantifies  the ``likeness'' of a given $N$-particle state $|\psi\rangle$ 
to that point, so that the set of these squared projections for all 
$\vartheta,\varphi$, referred to as a Husimi distribution, provides a 
phase-space representation of $|\psi\rangle$. 

We now investigate the Husimi distributions of the $N$-particle Floquet states 
of the periodically driven Bose-Hubbard dimer~(\ref{eq:PDD}). For the sake of  
graphical and computational convenience here we do not resort to the full 
time-dependent Floquet states~(\ref{eq:FFS}), but concentrate on their 
fixed-time intersections $| \psi_n(t=0) \rangle = | n \rangle$ with the
Poincar\'e plane. The Husimi portraits discussed in the following are obtained
by  superimposing the resulting distributions 
\be
	Q^{(N)}_{|n\rangle} (p,\varphi) = 
	\big| \langle n | \vartheta,\varphi \rangle_{\!N} \big|^2  
\label{eq:PSP}	
\ee
onto the corresponding mean-field Poincar\'e section already depicted in 
Fig.~\ref{F_1}, encoded in colors such that brighter colors indicate larger 
overlaps.

\begin{figure}[t]
\centering
\includegraphics[width=1.0\linewidth]{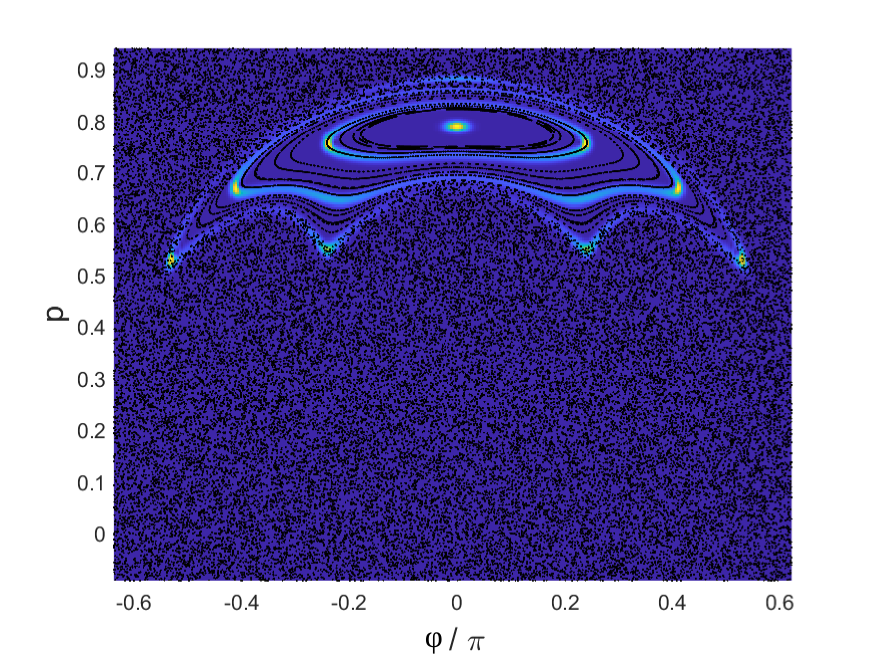}
\caption{Color-coded Husimi distributions~(\ref{eq:PSP}) of four selected 
Floquet states for $N = 10\,000$ Bose particles, labeled according to their 
degree of simplicity~(\ref{eq:DOS}) as
$n = 9520$ with $\eta_n = 0.274$, 
$n = 9700$ with $\eta_n = 0.695$, 
$n = 9900$ with $\eta_n = 0.887$,  and 
$n = 10\,000$ with $\eta_n = 0.996$ (outer to inner), superimposed on 
a magnified part of the Poincar\'e section displayed in Fig.~\ref{F_1}.}
\label{F_2}
\end{figure}

In Fig.~\ref{F_2}, we show such Husimi portraits of four particular Floquet 
states of a periodically driven Bose-Hubbard dimer filled with $N = 10\,000$ 
Bose particles; system para\-meters are the same as employed in Fig.~\ref{F_1}. 
Evidently these four states cling closely to the classical manifolds showing 
up in the island of regular motion, one of them being tied to the immediate 
vicinity of the stable elliptic fixed point,  two being  attached to closed 
curves surrounding this fixed point, and one located at the boundary to the 
stochastic sea. When including the time coordinate, the closed curves 
represent intersections with the Poincar\' e plane of  tubes $\mathbb T^+$ 
embedded in the odd-dimensional time-augmented phase space $(p,\varphi,\tau)$ 
which remain invariant under the Hamiltonian flow, while the elliptic fixed 
point marks the intersection of a stable periodic mean-field orbit with that 
plane. The full, time-dependent Floquet states~(\ref{eq:FFS}) then remain 
likewise attached to their respective, twisting invariant tube at each instant 
of time.

The association of $N$-particle Floquet states with these invariant 
mean-field tubes~$\mathbb T^+$ embedded in  the time-augmented phase 
space~\cite{BreuerHolthaus91} is analogous to the asymptotic connection 
between energy eigenstates of time-independent quantum systems possessing an 
integrable classical counterpart and the invariant tori in the even-dimensional 
conventional phase spaces of the latter, as specified by the semiclassical 
Einstein-Brillouin-Keller (EBK) quantization procedure~\cite{Gutzwiller90,
Haake10,Keller58,KellerRubinow60}. Namely, those discrete 
tubes~$\mathbb T^+\!(k)$ that ``carry'' a Floquet state are selected by the 
Bohr-Sommerfeld-like condition
\be
	\frac{1}{2\pi} \oint_{\gamma_k} \!\!\! p \, \rd \varphi \stackrel{!}{=} 
	\hbar_{\rm eff}\!\left( k + \frac{1}{2} \right) \; ,
\label{eq:BSC}
\ee
where $k = 0,1,2,\ldots$ is an integer quantum number, and $\gamma_k$ is a path 
winding once around the tube thus determined; this path can be continuously 
deformed such that it falls fully into the Poincar\'e plane. A second type 
of path following $\mathbb T^+\!(k)$ in time then provides a semi\-classical 
approximation to the quasienergy $\varepsilon_k$~\cite{BreuerHolthaus91,
SeligmannEtAl25}. As a consequence of  the scaling~(\ref{eq:SCA}) and the
definition~(\ref{eq:PIM}) of the momentum~$p$ the effective Planck constant 
appearing on the right-hand side of this condition~(\ref{eq:BSC}) is given by
\be
	\hbar_{\rm eff} = \frac{2}{N} \; .
\label{eq:EPC}
\ee
Hence, the particle number~$N$ determines the scale at which the 
quantum dynamics can adhere to the mean-field prediction: The larger~$N$, 
the smaller is $\hbar_{\rm eff}$, and the finer are the details of the 
mean-field phase space that the quantum $N$-particle system is able to 
resolve. Of course, this is just another view on the supposed approach of the 
quantum system to its mean-field limit with increasing~$N$, while $\alpha$
is kept constant.

The link between the exact $N$-particle Floquet states and the exact 
mean-field tubes manifesting itself in Fig.~2 convincingly confirms these
semiclassical considerations. In particular, the Floquet state most compactly
concentrated around the stable periodic orbit clings to the innermost 
quantized tube $\mathbb T^+\!(0)$ in the manner specified by 
Eq.~(\ref{eq:BSC}), thus representing an effective ground state $k = 0$ of 
the resonant regular island, whereas the other Floquet states that are 
semiclassically associated with wider tubes in that island represent excited 
states $k > 0$. In this way, condition~(\ref{eq:BSC}) automatically 
induces an ordering at least of these ``regular'' Floquet states.

This semiclassical perspective obviously reverses our initial line of thought:
First, the actual $N$-particle quantum dynamics had been subjected to 
a mean-field approximation, leading to the classical-like Gross-Pitaevskii 
equation~(\ref{eq:GPE}) or its pendulum equivalent~(\ref{eq:EQF}), whereas
now the regular part of the classical dynamics has been ``requantized''  by 
means of Eq.~(\ref{eq:BSC}). Notwithstanding the conceptual insight gained
in this manner, the Gross-Pitaevskii equation effectively is a single-particle 
equation for a Hartree-type product state~\cite{Leggett01}, so that the above 
procedure can be consistent only if the exact $N$-particle Floquet states are 
``simple'' in the sense that they constitute macroscopically occupied, 
periodically time-dependent single-particle orbitals, {\em i.e.\/}, periodically 
driven Bose-Einstein condensates that evolve in time without heating. 
Therefore,  one also needs a tool that allows one to decide whether or not 
this is the  case.

As one such tool, we suggest the ``degree of simplicity'' (coherence) discussed
by Leggett in the context of time-independent Bose-Einstein     
condensates~\cite{Leggett01}. When applied to the periodically driven 
Bose-Hubbard dimer~(\ref{eq:PDD}), this prompts us to compute the one-particle
reduced density matrices
\begin{equation}
	\varrho_n = \left( \begin{array}{cc}
	\langle \ad_1 \ab_1 \rangle_n & \langle \ad_1 \ab_2 \rangle_n \\
	\langle \ad_2 \ab_1 \rangle_n & \langle \ad_2 \ab_2 \rangle_n 
		\end{array} \right) \; , 
\end{equation}		
where $\langle \ad_j \ab_k \rangle_n$ denote  the expectation values
$\langle n | \ad_j \ab_k | n \rangle$ taken with the eigenvectors $|n\rangle$ 
of the one-cycle time evolution operator~(\ref{eq:DIA}), and to evaluate the
quantities
\begin{equation}
	\eta_n  = 2 N^{-2} \, {\rm tr} \, \varrho_n^2 - 1 \; . 
\label{eq:DOS}
\end{equation}	
A main virtue of this construction lies in the fact that the trace of a matrix
does not depend on its basis, so that this degree of simplicity~(\ref{eq:DOS}) 
can be obtained without knowledge of the macroscopically occupied state, if 
there is one. Evidently, an eigenvector $|n\rangle$ that equals an $N$-fold 
occupied  single-particle state yields $\eta_n = 1$, whereas a maximally 
fragmented state gives $\eta_n = 0$. Moreover, while there is no order of
the Floquet states with respect to their quasienergy, they can conveniently be 
ordered according to the magnitude of~$\eta_n$. 

\begin{figure}[t]
\centering
\includegraphics[width=1.0\linewidth]{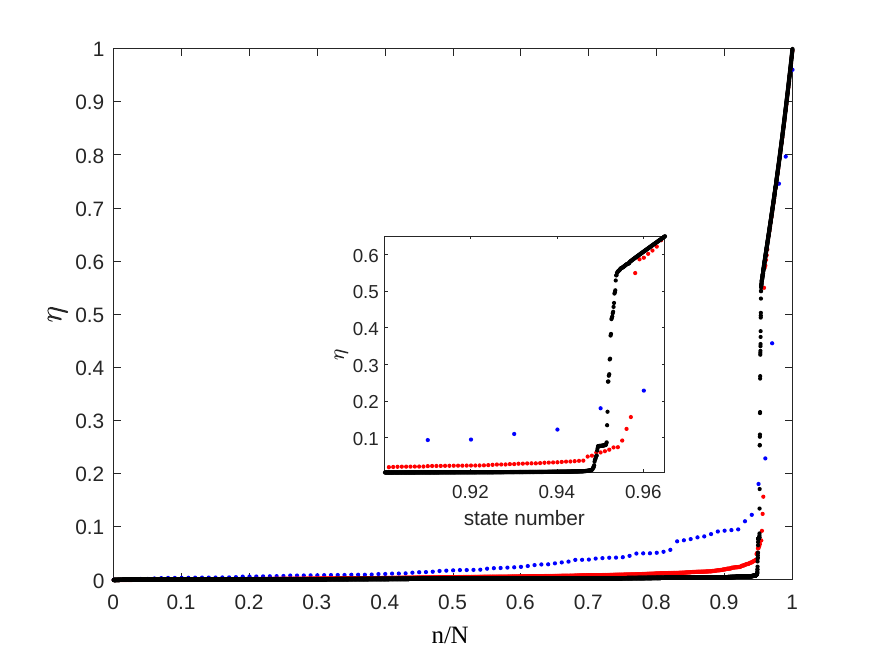}
\caption{Degree of coherence~(\ref{eq:DOS}) for the Floquet states of the 
periodically driven Bose-Hubbard dimer with parameters as used before in 
Figs.~\ref{F_1} and~\ref{F_2}, for particle numbers $N = 100$ (blue), 
$N = 1000$ (red), and $N = 10\,000$ (black). State labels~$n$ have been assigned 
in accordance with the magnitude of~$\eta_n$, with the scaled state number
$n/N$ appearing on the abscissa.}
\label{F_3}
\end{figure}

Figure~\ref{F_3} depicts the degree of simplicity for all Floquet states of
the periodically driven Bose-Hubbard dimer with parameters as before, for
$N = 100$,  $N = 1000$, and $N = 10\,000$. Here, the Floquet states have been
ordered such that their state label~$n$ increases with increasing~$\eta_n$,
beginning with $n = 0$. In order to compare data for different~$N$, 
the degree is plotted vs the scaled state number~$n/N$. While the curves 
connecting the data points for $N = 100$ or $N = 1000$ still change 
appreciably when the particle number is increased further, data obtained for 
$N=10\,000$ appear to be stable at the level of graphical resolution, apart 
from the appearance of some fine structure that becomes visible under
 the increased $\hbar_{\rm eff}$-controlled resolution, such as the step 
clearly visible in the inset. The resulting (black) curve exhibits a 
fairly sharp kink at about $n/N \approx 0.95$, indicating a  transition from 
complex, {\em i.e.\/}, nonsimple Floquet states with degrees~(\ref{eq:DOS}) 
close to zero to more coherent states. This is the transition from the Floquet 
states associated with the chaotic sea to the Floquet states associated with 
the main regular island. As examples for the latter, the Floquet states 
previously shown in Fig.~\ref{F_2} carry the simplicity-ordered labels 
$n = 9520$, $9700$, $9900$, and $10\,000$, respectively, yielding 
$\eta_{9520} = 0.274$, $\eta_{9700} = 0.695$, $\eta_{9900} = 0.887$, and 
$\eta_{10\,000} = 0.996$. A comparison of this $\eta$-induced order with the 
$k$-induced order implied by Eq.~(\ref{eq:BSC}) now reveals a key feature: 
Within the regular island, the ordering of the Floquet states with respect to 
their degree of simplicity precisely matches the ordering with respect to the 
semiclassical quantum number~$k$ introduced in the Bohr-Sommerfeld-like 
condition~(\ref{eq:BSC}), such that states associated with successively 
narrower quantized invariant tubes, that is, with successively lower~$k$, 
possess a successively increasing degree of simplicity. In short, one has
$k = N - n$ for sufficiently large~$N$ and regular island states. The state 
$k =0$ or $n = N$ associated with the innermost quantized tube at the top 
of this $\eta$-hier\-archy is almost fully coherent, as witnessed by its 
degree of simplicity close to unity. This observation implies that for initial 
conditions placed in the center of the resonant regular island the 
Gross-Pitaevskii equation will provide a trustworthy image of the $N$-particle 
evolution for long times.
  
\begin{figure}[t]
\centering
\includegraphics[width=1.0\linewidth]{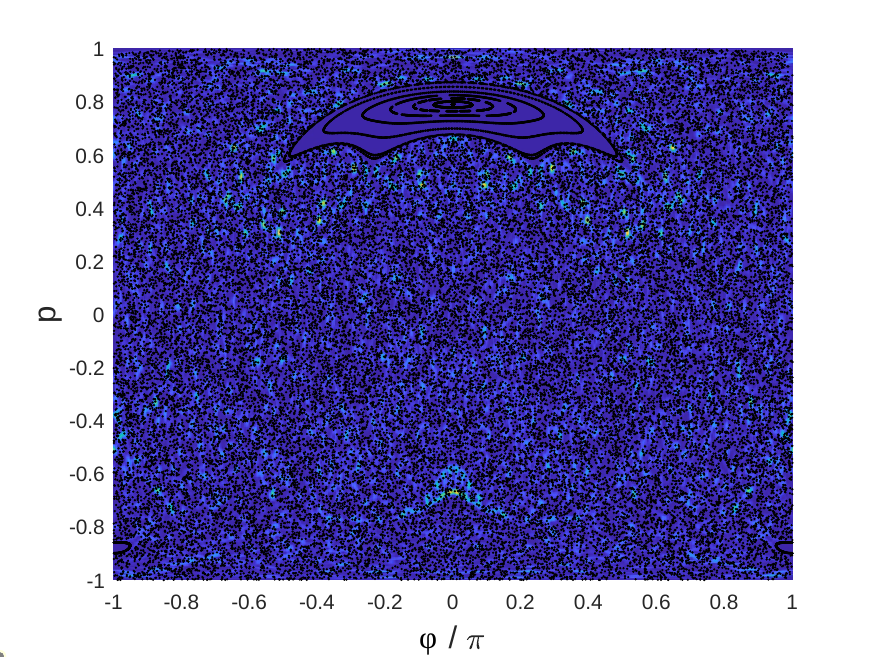}
\caption{Color-coded Husimi distribution of the Floquet state
 for $N = 10\,000$ with the lowest degree of coherence , $n = 0$ 
 with  $\eta_0 = 6.7 \times 10^{-7}$.}     
\label{F_4}
\end{figure}

At the other end of the $\eta$-scale, we display in Fig.~\ref{F_4} the Husimi 
plot of the most complex, noncoherent Floquet state $n = 0$ showing up
for the present para\-meters.  This state appears to be mainly associated with
the chaotic sea, albeit not in a homogeneous manner. Instead, one observes a 
speckle pattern predominantly concentrated at some distance from the 
regular island. For initial conditions exhausted by states of this kind, the 
time-dependent Gross-Pitaevskii equation cannot be expected to capture the 
actual $N$-particle quantum dynamics. It needs to be kept in mind, though, 
that the computation of these 
Floquet states is plagued by the almost degeneracy of their quasienergies, 
so that inevitable numerical inaccuracies may tend to hybridize states with 
too closely spaced eigenvalues. Note that this seemingly artificial feature 
actually does possess a physical analog, since even tiny perturbations would 
cause transitions between such states. 

\begin{figure}[t]
\centering
\includegraphics[width=1.0\linewidth]{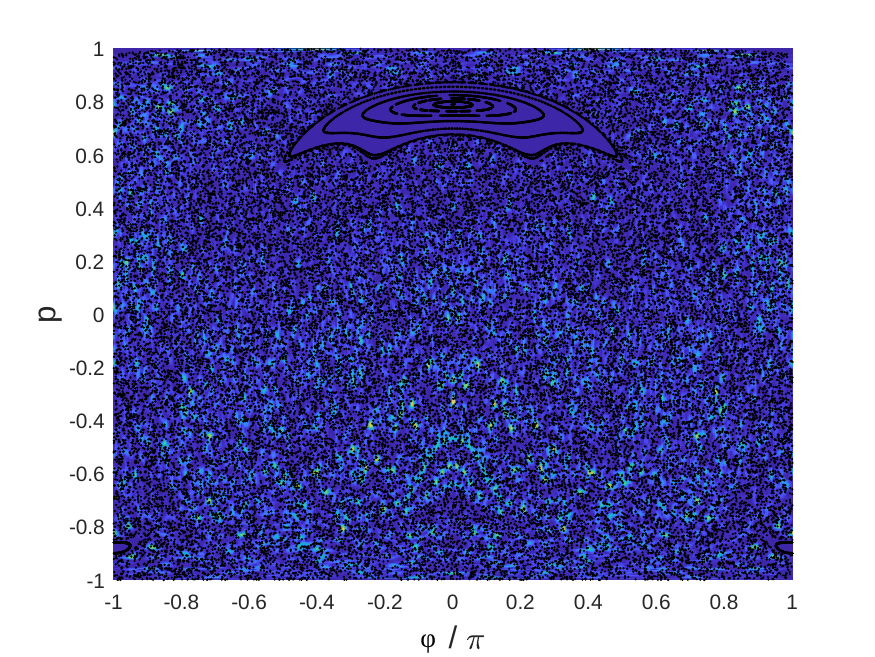}
\caption{Color-coded Husimi distribution of a Floquet state for 
$N = 10\,000$  falling into the transition regime, labeled $n = 9479$ 
with $\eta_{9479} = 0.0117$.}     
\label{F_5}
\end{figure}

\begin{figure}[t]
\centering
\includegraphics[width=1.0\linewidth]{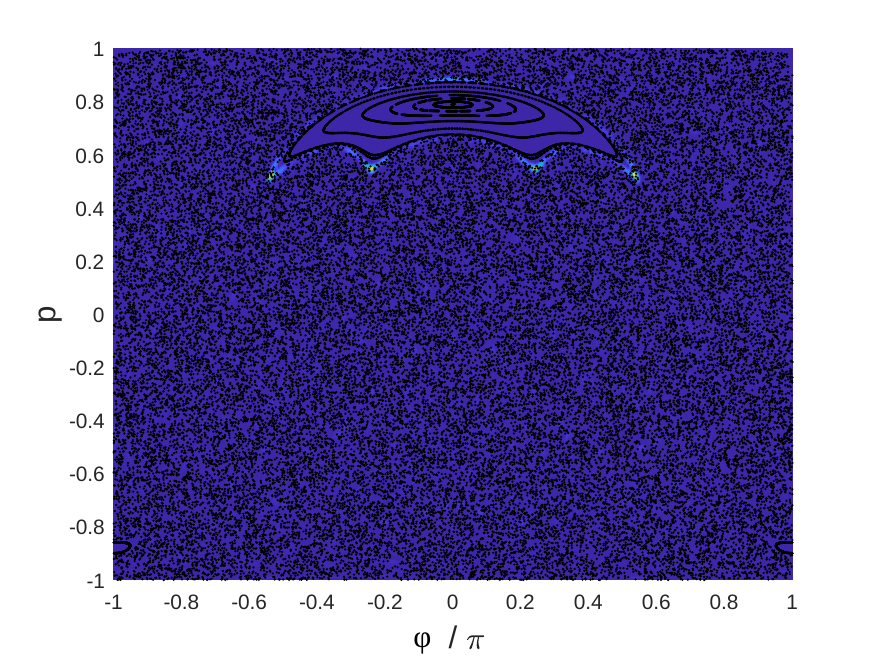}
\caption{Color-coded Husimi distribution of the Floquet state for 
$N = 10\,000$ labeled $n = 9480$ with $\eta_{9480} = 0.0135$.
This is a transitional state between the ``chaotic'' and the ``regular'' ones,
with its label~$n$ exceeding that of the state displayed in the previous 
Fig.~\ref{F_5} by only $\Delta n = 1$.}     
\label{F_6}
\end{figure}

As already remarked above, the system parameters employed throughout this 
work have been chosen such that the transition from stochastic to regular 
motion appears to be particularly sharp on the mean-field level. This 
sharpness is reflected in the Floquet states, as demonstrated by a comparison 
of  Figs.~\ref{F_5} and~\ref{F_6}, which exhibit two states neighboring with 
respect to their simplicity within the transition regime, $n = 9479$ and 
$n =  9480$. While the former is still spread widely over the chaotic sea, 
the latter is already concentrated along the borderline of the regular island, 
similar to the state $n = 9520$ portrayed in Fig.~\ref{F_2}. 

Again there is a caveat. Although the boundary that separates the chaotic 
sea from the regular island appears to be fairly sharp on the scale of 
resolution adopted in Fig.~\ref{F_1}, there actually is some small, intricate 
fine structure, as implied by the general theory of Hamiltonian 
systems~\cite{AbrahamMarsden08,LiLi92,GuckenheimerHolmes02}. This fine 
structure would gradually be resolved by the $N$-particle system when $N$ is 
increased further while keeping the mean-field parameter constant, possibly 
leading to signatures in an $\eta$-plot similar to the small step visible in 
the inset of Fig.~\ref{F_2}. The detailed implications of this scenario for 
the degree of simplicity of the participating Floquet states merit further 
analysis.

\section{Conclusions and outlook}
\label{S_4}

The cornerstones of our discussion may be summarized as follows:  
The periodically time-dependent Gross-Pitaevskii equation, as a nonlinear 
mean-field approximation to the Schr\"odinger equation of a periodically 
driven, interacting $N$-particle system governed by a Hamiltonian 
$H(t) = H(t+T)$, naturally possesses stable $T$-periodic orbits with 
surrounding zones of regular, {\em i.e.\/}, nonchaotic dynamics. When the 
particle number~$N$ is sufficiently large, quantizable tubes in the sense of 
the condition~(\ref{eq:BSC}) fit into these zones. The innermost of these 
tubes provides the backbone for the EBK-like construction of a fully coherent, 
``simple'' many-body Floquet state, which constitutes a periodically 
time-dependent, $N$-fold occupied single-particle orbital, facilitating a 
periodically time-dependent Bose-Einstein condensate that evolves without 
heating on the mean-field level.

In this context there is a further subtle issue that deserves to 
be mentioned explicitly. Namely, the unresolved fine-scale chaos within a 
mainly regular mean-field island also possesses a quantum counterpart, giving 
rise to tiny avoided crossings between quasienergies belonging to ``regular'' 
and ``chaotic'' Floquet states. Although these may be, once again, far too 
narrow to be computationally detectable for truly large~$N$ they must, as a 
matter of  principle, exist~\cite{Haake10}, effectuating quantum mechanical 
tunneling between both types of states. Clearly, this is a beyond-mean-field 
effect which would ultimately render a Floquet condensate metastable, on 
timescales inversely proportional to the widths of such hyperfine avoided 
crossings. We speculate that those timescales would not be relevant for 
practical purposes, but so far these qualitative considerations have not been 
turned into quantitative estimates.   

While the above interrelationships have been exemplified here with the help of 
a model system which has been deliberately kept so elementary that each link 
in the chain of arguments could be verified by comparison with the exact 
Floquet solutions of its $N$-particle Schr\"odinger equation, we surmise that 
they also hold for more sophisticated, experimentally accessible arrangements 
for which the  related Gross-Pitaevskii equation can still be solved 
numerically, whereas the solution of the full Schr\"odinger equation will not 
be feasible even with most powerful future supercomputers. Nonetheless, the 
present case study suggests that periodic solutions of the Gross-Pitaevskii 
equation may allow one to reliably predict conditions under which the 
envisioned Floquet condensates do exist.  Further investigations along these 
lines, both theoretical and experimental, should shed additional light on this 
subject. To this end, an experimental setting which appears to be 
particularly promising is a periodically driven optical lattice onto which 
an additional harmonic potential created by external electromagnets is 
superimposed, as recently reported in Ref.~\cite{CaoEtAl20}. While a linear 
external field would lead to a Wannier-Stark ladder of equidistant energies, 
the harmonic trap induces a slow variation of the spacings of the on-site 
energies, as corresponding to the anharmonic pendulum motion. Indeed, 
the authors of Ref.~\cite{CaoEtAl20} already have discussed, for a specific 
driving scheme, the implications of this effective anharmonicity on the 
classical phase-space diagram, and have been able to measure phase-dependent 
spatial dynamics. Such experimental developments may come fairly close to what 
is needed for verification of the ideas put forward in the present work.

\section*{Acknowledgments}
This work has been supported by the Deutsche Forschungsgemeinschaft (DFG, 
German Research Foundation) through Project No.~355031190.  We thank the
members of the research group FOR2692 for inspiring discussions.

\end{document}